\documentstyle[12pt]{article}
\begin{document}
\setlength{\baselineskip}{3ex}
\begin{center}
{\large\bf Coalescence, Percolation and Nuclear Multifragmentation}\\
\vspace{1cm}
S. Das Gupta, C. Gale and K. Haglin \\
{\em Physics Dept., McGill University, Montreal, P.Q., H3A 2T8, Canada.}\\
\vfill
{\bf Abstract}
\end{center}
\setlength{\baselineskip}{5ex}
We show that the coalescence model for fragment formation leads to an
approximate site percolation model.  Features characteristic of a percolation
model also appear in microscopic models of disassembly.
\vfill
--------------------------- 

\pagebreak

Multifragmentation continues to be one of the most important aspects of
heavy ion collisions.  Speculations have been made
that the observed distribution of mass fragments may indicate that
during disassembly finite amounts of nuclear matter passed
through a critical region of liquid-gas phase transition$^1$. On the other
hand, there have been
practical calculations which directly attempt to calculate mass distributions
using different models which typically have only indirect links, if any,
to the question  of phase transition.  For example, there is a modern
version of the evaporation model$^2$, statistical models both simultaneous
and sequential (with and without radial expansion built in)$^{3-5}$,
and a fully microscopic model proposed in ref. 6
which is based upon the BUU model but with the inclusion of fluctuations.
Applications of this BUU-with-fluctuations$^{7,8}$ model
agree reasonably well with experimental
data but these calculations are very computer intensive so it is difficult
to extract simple physics from such calculations.  However, a
critical study of the validity of this model has been made$^9$.

The percolation model was introduced by Bauer$^{10}$ et. al. and
Campi et.al.$^{11}$ in an attempt to model nuclear fragmentation.
Numerous applications to
data have been made$^{12}$ and Campi has argued that one can use this model
to find out how, in spite of of finite particle number, signatures of phase
transitions can be extracted from experiments$^{13}$.  Both Campi and Bauer
have provided some physical arguments for the parameters of percolation
models.  For bond percolation model in three dimensions,
the six bonds refer to the attractive
bonds a given nucleon will feel because of short-range attractive nuclear
interaction$^{12}$.  These bonds are then broken with a probability $p$, which
is the percolation parameter and depends linearly on the excitation energy
per nucleon.  While this picture is intuitively very pleasing and easily
understandable for bonds between atoms in a molecule, for example, it is
harder to understand it in terms of mean-field theories of nuclei (Hartree-Fock
or Thomas-Fermi) which are usually employed to describe finite nuclei.  In
this paper we attempt to overcome this difficulty.
We will show that a  simple coalescence model for fragment formation
leads to an approximate site percolation model in three dimensions.
Although the arguments presented for the connection between the two models
are based on many approximations they seem to hold nonetheless as substantiated
by results of a mean field calculation that we will present at the end.  That
calculation as well establishes contact between percolation and microscopic
models.

A popular version of the coalescence model is the following:
If, after hard collisions are over, $n$ nucleons appear within a short momentum
distance $P_0$ of each other then these nucleons will coalesce into one
composite because of mutual attractive interactions.  This simple idea is
incomplete.  Two nucleons with about the same momenta might be spatially
separated by a large distance clearly prohibiting them from coalescing
to form a deuteron.  The real statement should be that nucleons
must appear close to each other in phase-space$^{15}$.  On a semi-classical
level the volume of a cell in phase-space is $h^3$.  In any one of these
phase-space cells at most one nucleon with a given spin-isospin can appear.
We will assume for simplicity our system has $N=Z$ and no spin excess.  After
hard collisions are over nucleons will then find themselves in the
six-dimensional space which we partition into cubes each of volume $h^3$.  A
given cube then has six labels $i,j,k,l,m,n$ where the last three labels
refer to momenta and the first three to configuration space.  A given cube
can accommodate more than one nucleon (up to four) but then they must belong
to different spin-isospin classes.
We might argue that nucleons in the same cell and in adjoining cells with
one common wall form a cluster; different clusters have empty walls between
them.  This leads to a
site percolation model in six dimensions.  However, the physics of the problem
leads to a simpler situation.

We have stated that the volume of each phase-space cell is
$(\Delta r)^3(\Delta p)^3=h^3$ but we have not set the relative length scales
for $\Delta r$ or $\Delta p$.  We can choose them to our advantage.
A remarkable feature of nuclear physics is that as the mass number $A$
grows so does the volume in configuration space but the volume in momentum
space remains roughly constant; it is always given by $\frac {4\pi }{3}p_F^3$
where $p_F$ is independent of the mass number $A$.  Hence we
choose $\Delta p$ from $(\Delta p)^3=\frac {4\pi }{3}p_F^3$.  For clusters
to form, the nucleons must belong to the same momentum cell (otherwise
there is far too much momentum in the system for it to stay together) although
they can be in adjoining configuration space cells.  Thus for each
value $l,m$ and $n$ of momentum indices, we have a
site percolation model in three dimensions in configuration space.  Refs. 12
and 13 use a bond percolation model in three dimensions; for problems in
nuclear physics the bond percolation bond is easier to use
as the number of nucleons is always fixed independent of
how many bonds are open or closed.  The situation is more complicated for
site percolation; one assigns an occupation probability
$p$ and, if the number of lattice sites is $N$, then the average number of
nucleons is $Np$.  Thus if the value of $p$ changes, $N$ must also change
if we want to model the same nucleus.  Since most studies use Monte-Carlo
simulations the number of occupied sites will usually fluctuate around the
value $Np$ unless special care is taken.  Nonetheless, in the spirit of
drawing an analogy between percolation and coalescence as discussed
above, the parameters for site percolation take physical meaning.  The
parameter $p$ becomes related to entropy which can be computed in numerical
simulations as done by Bertsch and Cugnon$^{16}$. We also see that $p$ can
vary with position in momentum
space, for example, $p$ should drop as the value of the momentum
characterizing a momentum cell increases.  Finally $p$ can also vary
with position in configuration space.  The increase in the value of $N$ as
$p$ changes merely reflects the fact that more and more of phase-space is
becoming accessible as, for example, will happen if the energy of
collision increases.  We remind the reader that because of
universality,$^{17}$ the conclusions reached by the use of site
percolation model remain unchanged.

The Pauli principle plays a crucial role in the equivalence
between percolation and coalescence. This may seem surprising at first
glance but we remind the reader that the Pauli principle was
always respected in other studies of phase-transitions in nuclear matter
at intermediate energies.  For example, in the mean field studies of ref. 18
a given quantum cell does not contain more than one particle of a given
spin-isospin.

We still need to show that the simple coalescence model
with the incorporation of spin-isospin can be mapped quite faithfully into
another calculation which uses no such indices.  Let the first calculation
which explicitly uses spin-isospin have $N$ lattice sites and occupation
probability $p$ for each spin-isospin species.  Here each lattice site
may have either
no nucleon or between 1 and 4 nucleons.  In a corresponding calculation
without explicit recognition of spin-isospin let the number of lattice sites
be $N'$ and $p'$.  Here each lattice site has either no nucleon or 1 nucleon.
What is the mapping of $N'$
to $N$ and $p'$ to $p$?

First of all we want to describe the same nucleus which requires
\begin{eqnarray}
N'p'=4Np
\end{eqnarray}
Secondly, the physics of the problem is dictated largely by the ratio of
the number of occupied sites to the number of unoccupied sites; we want these
ratios to be the same in both the descriptions; this gives
\begin{eqnarray}
1-p'=(1-p)^4
\end{eqnarray}
Together these two equations determine $N$ and $p$ in terms of $N'$ and $p'$.

Figs. 1 and 2 show that the same physics is described whether or not
spin-isospin is explicitly included and also shows the validity of the
mapping described by eqs. (1) and (2).  Here we have used the site
percolation model.  Fig. 1 is the
``Campi" plot$^{13}$.  We choose arbitrarily the total number of nucleons
to be $A=64$; for a given choice of $N$, $p$ is given by $A/4N$; similarly, for
a given choice of $N'$, $p'$ is given by $A/N'$.  Monte-Carlo simulation
produces $A'$ particles where $A'$ fluctuates around $A$.  We define reduced
multiplicity $n$ as the number of clusters divided by  $A'$.  For a given
$p(p')$ the quantity $n$ will vary; nonetheless an average $<n>$ can be defined
for each $p(p')$ and in figs. 1 and 2 when plotting against $<n>$ we actually
include events differing from $<n>$ by at most $\pm .03$.  For a theoretical
calculation, it is easier to use the variable $p(p')$, but $p(p')$ is not
an observable whereas $n$ is a direct observable.
Each point in Figs. 1 and 2
are averages over 1000 runs taken for fixed $N(N'),p(p')$ and the small
dispersion about $<n>$.
The second moment $M_2$ is defined as $M_2=\sum\limits_{s} s^2d(s)/A$ where
$d(s)$ is the average value of the number of clusters
of size $s$ and the largest cluster in each event is left out when this
averaging is computed.  The definitions follow those of ref. 13.
Fig. 2 shows the average value of the mass number
of the largest cluster once again as a function of $n$ in order to show
that results with explicit spin-isospin can be mapped onto results
without explicit spin-isospin.  We have shown in Figs. 1 and 2 results
from two-dimensional site percolation rather than three in order to gather
more points ($N=L^3$ grows
very quickly with $L$ so that the number of points in the interesting region
will be sparse).  Three dimensional percolation model calculations were also
done; with fewer available points they confirm the conclusions presented
here.

The simple coalescence model with which we have established a one-to-one
correspondence to the percolation model is not sufficiently realistic.
One may wonder if features seen in percolation model can emerge in more
realistic microscopic models which nonetheless depend upon particles being
near each other in phase-space coalescing to form clusters.  We will follow
numerically the time development of a blob of nuclear matter which is
initially at excitation energy and compression appropriate for about
40 MeV/nucleon beam energy for equal ion collisions.
We assume that because of hard collisions a phase-space density with
fluctuations has been produced which then disassembles; during disassembly
matter is moving outward so there are fewer collisions.  We take into
account only a mean-field during the disassembly.  This model was first
employed by Knoll and Strack$^{19}$ who used time-dependent Hartree-Fock theory
in two dimensions.  The model here is identical with that of ref. 14
except the initial stage in ref. 14 was produced by the cascade model. At
the initial time, the $i-$th nucleon is represented by a phase-space density
\begin{eqnarray}
f(\vec r,\vec p)=\frac {1}{(\hbar \pi )^3}exp(-\alpha ^2(\vec r-\vec r_i)^2)
exp(-(\vec p-\vec p_i)^2)/\hbar ^2\alpha ^2).
\end{eqnarray}
This is a Wigner transform of a Gaussian wave-function and retains the
quantum uncertainty between $\vec r$ and $\vec p$, i.e., if the nucleon is
localized in configuration space, then it is spread out in momentum space.
The value of $\alpha $ is kept at 0.5$fm^{-1}$ which is a standard value used
before in similar calculations$^{14,20}$.  We choose the initial positions
$\vec r_i$ by Monte-Carlo sampling within a radius $R$ which is fixed
such that the average
density $\bar{\rho }=\int \rho ^2d^3r/\int \rho d^3r$ is $0.18fm^{-3}$.  The
initial momenta $\vec p_i$ are chosen by Monte-Carlo sampling an
occupational probability distribution obtained from a nuclear matter
Fermi-gas calculation at density
$.18fm^{-3}$ and a given temperature.  Small adjustments (detailed later) are
made so that the initial energy of the generated phase-space density has a
pre-assigned value.  Ideally in subsequent Vlasov propagation this energy is
exactly conserved.  We use the test particle method and the Lenk-Pandharipande
prescription$^{21}$ which gives quite accurate energy conservation.  We use
50 test particles per nucleon.  The calculation is done in a configuration
space
$(40 fm)^3$ and the size of each cube is $1fm^3$. A Skyrme interaction
$U(\rho )=A(\rho /\rho _0)+B(\rho/\rho _0)^{7/6}$ is used where $A=-356 MeV,
B=303 MeV$ and $\rho _0=.16fm^{-3}$.
We start with a small and compact piece of
nuclear matter and let it evolve for $81 fm/c$.  At this
time we count clusters.  Test particles which share a common wall are part of
the same cluster; the number of real particles in each cluster will usually
turn
out to be non-integral.  We disregard initially all the clusters whose number
of particles adds up to less than one; they will contribute to single nucleons.
The clusters with particle number greater than one are integerized to the next
integer.  The number of single nucleons is then the total number of nucleons
minus the number of nucleons bound in composites.   One could also
let the system evolve for longer times but this would necessitate expanding
the dimension beyond $(40 fm)^3$ as some nucleons begin to leave the box.

  The results of our calculation are shown in Fig. 3.  We plot the second
moment
$<M_2>$ as function of energy per nucleon rather than $<n>$ since in the
calculation the energy can be specified at the beginning whereas $n$ is
determined only at the end when Vlasov propagation stops.  We find that
$<M_2(n)>=f(n,E/A)$ and thus for a meaningful plot against $n$ we need to
choose a weighting function $g(E/A)$.  There is of course an average value
$<n>$ for a given $E/A$ which is shown also in Fig.3. If we plotted $<M_2>$
against the calculated $<n>$ the graph would look very much like Fig.2 in
ref. 13.  In Fig. 3 we have shown the results for systems with particle
numbers 40 and 64.  Not only does one find the appearance of a maximum, the
height of this maximum increases as the number of particles in the system
increases much like in the percolation model in 3 dimensions.  The value of
$E/A$ can
be converted into a beam energy remembering that with our forces and
approximations the energy per nucleon in a cold nucleus of 40 nucleons is
$\approx -11 MeV$.  It is probably not surprising that the maximum in $M_2$
is obtained
when the available beam energy is close to that which is enough to liberate
all the nucleons.  Each point in Fig. 3 is an average of 40 runs at a
given $E/A$.

Although at initialization the momenta in $\vec p_i$ (eq.3) were chosen from
a nuclear matter Fermi gas calculation at a given temperature, neither the
kinetic energy nor the potential energy per particle in our system
will be the same as in nuclear matter.  The potential
energy is different because in our system the density fluctuates; the kinetic
energy per particle is also different because the folding done in eq.3
increases the kinetic energy. In addition Monte-Carlo sampling leads to
usual small fluctuations in the value of $E/A$ from one event to another.
In the numerical simulation
once the Monte-Carlo for positions and momenta of test particles is done for
one event a slight readjustment of the momentum scale
is imposed so that we always have a fixed energy per particle.

It is possible to see that the appearance of a maximum in Fig. 3 (or Fig. 2 in
ref. 13) is closely related to composite formation.  Imagine a scenario
where a piece of hot nuclear matter containing $A$ nucleons
can only shed nucleons but not composites.
If the excitation energy is small it sheds none or a few nucleons.  On the
other extreme for high excitation energy it will explode into $A$ nucleons.
In the low excitation energy limit $n$ is $1/A\approx 0$ and $M_2= 0$
since the largest cluster is left out when computing $M_2$.  On the high
energy side $n\approx 1$ and $M_2\approx 1$.  The value of $M_2$ grows
monotonically from 0 in one extreme to 1 in the other; thus the appearance
of a maximum in the plot of $M_2$ against $n$ and a value higher than 1
can only happen if fragments are formed.

It will be interesting to check whether the features seen in figures 1 and 3
also appear in phenomenological models such as considered in ref. 22.

Our interest in the present problem grew as a result of discussions with
Dr. Campi while he was visiting McGill University.  We wish to thank Dr.
Campi and Dr. de Takacsy for discussions.

\centerline{REFERENCES}

1)  M. W. Curtin, H. Toki and D. K. Scott, Phys. Lett. B123(1983)289

2)  W. A. Friedman, Ann. Phys. (N. Y.)192(1989)146

3)  D. H. Gross, Nucl. Phys. A488(1988)217c

4)  J. P. Bondorf, R. Donangelo, H. Schulz and K. Sneppen, Phys. Lett. B162
(1985)30

5)  G. Fai and J. Randrup, Nucl. Phys. A404(1983)551

6)  W. Bauer, G. F. Bertsch and S. Das Gupta, Phys. Rev. Lett. 58(1987)863

7)  J. Gallego, S. Das Gupta, C. Gale, S. J. Lee, C. Pruneau and S. Gilbert,
Phys. Rev C44(1991)463

8)  H. H. Gan, S. J. Lee and S. Das Gupta,Phys Rev. C36(1987)2365

9)  Ph. Chomaz, G. F. Burgio and J. Randrup, Phys. Lett B254(1991)340

10) W. Bauer, D. R. Dean, U. Mosel and U. Post, Proceedings of the seventh
high energy heavy ion study, GSI Darmstadt, 1984, GSI Report-85-10,pp 701

11) X. Campi and J. Debois, Proceedings of the seventh high energy heavy ion
study, GSI Darmstadt, 1984, GSI Report-85-10,pp 707

12) W. Bauer, Phys. Rev C38(1988)1297

13) X. Campi, Phys. Lett B208(1988)351

14) C. Gale and S. Das Gupta, Phys. Lett B162(1985)35

15) B. K. Jennings, S. Das Gupta and N. Mobed, Phys. Rev C25(1982)278

16) G. F. Bertsch and J. Cugnon, Phys. Rev C24(1981)2514

17) D. Stauffer, Introduction to percolation theory(Taylor and Francis, London,
1985)

18) H. Jaquaman, A. Z. Mekjian and L. Zamick, Phys. Rev C27(1983)2782

19) J. Knoll and B. Strack, Phys. Lett. B149(1984)45

20) D. H. Boal and J. N. Glosli, Phys Rev C38(1988)1870

21) R. Lenk and V. Pandharipande, Phys. Rev C39(1989)2242

22) S. J. Lee and A. Z. Mekjian, Phys. Rev. C45(1992)1284.

\vfill
\eject
\centerline{Figure Captions}

Fig.1. Plots of second moment against reduced multiplicity done without
explicit consideration of spin-isospin (solid dots) and with explicit
consideration of spin-isospin (hollow dots).  The bracketed numbers
opposite solid dots give the value of $p'$; the bracketed numbers opposite
hollow dots do not refer to $p$ but
to equivalent $p'=1-(1-p)^4$.  This curve therefore shows that the second
moment and multiplicity are mapped by eqs. (1) and (2).

Fig. 2. The average value of the mass number of the largest cluster
plotted against reduced multiplicity for the two calculations in Fig. 1.
The same horizontal axis is chosen to facilitate the
correspondence between points in Figs. 1 and 2.

Fig. 3. Second moment (left axis) as it depends on the energy per particle.
Also shown (on the right axis)
is the average reduced multiplicity against the same variable,
namely, the energy per particle.

\end{document}